\documentclass[12pt]{article}
\usepackage{epsfig,amsmath}

\parskip=\medskipamount 
\renewcommand{\thesection}{\arabic{section}}

\catcode`@=11

\@addtoreset{equation}{section}
\@addtoreset{equation}{subsection}
\def\theequation{\ifnum\value{section}=0 \arabic{equation}\ignorespaces
\else \ifnum\value{section}=-1 A.\arabic{equation}\ignorespaces
\else \ifnum\value{subsection}=0 \thesection.\arabic{equation}\ignorespaces
\else \thesection.\arabic{subsection}.\arabic{equation}\ignorespaces
                             \fi
                        \fi
                   \fi}

{\catcode`\'=\active \def'{{}^\bgroup\prim@s}}

\catcode`@=12

\newcommand{\bq}{\begin{equation}}
\newcommand{\be}{\begin{equation}} 
\newcommand{\fq}{\end{equation}}
\newcommand{\ee}{\end{equation}}
\newcommand{\bea}{\begin{eqnarray}}
\newcommand{\eea}{\end{eqnarray}}

\def\bop#1{\setbox0=\hbox{$#1M$}\mkern1.5mu
        \vbox{\hrule height0pt depth.04\ht0
        \hbox{\vrule width.04\ht0 height.9\ht0 \kern.9\ht0
        \vrule width.04\ht0}\hrule height.04\ht0}\mkern1.5mu}
   
\begin{document} 

\thispagestyle{empty}

\begin{flushright}
\begin{tabular}{l}
CTP-MIT-3157\\ 
 
\end{tabular}
\end{flushright}

\vskip.3in
\begin{center}
{\Large\bf Strong Coupling Phenomena \\ 
                on the Noncommutative Plane.}

\vskip.3in

\vskip .2in
{\bf Zachary Guralnik}
\\[5mm]
{\em Center for Theoretical Physics \\
Massachusetts Institute of Technology\\
Cambridge MA, 02139}\\
{Email: zack@mitlns.mit.edu}

\vskip.5in minus.2in

{\bf Abstract}

\end{center}

We study strong coupling phenomena in $U(1)$
gauge theory on the non- commutative plane. 
To do so, we make 
use of a T-dual description in terms of an $N\rightarrow\infty$ 
limit of $U(N)$ gauge theory on a commutative 
torus.  The magnetic flux on this torus is taken to be $m=N-1$,  while
the area scales like $1/N$,  keeping 
$\Lambda_{QCD}$ fixed.  
With a few assumptions, we argue that the   
speed of high frequency light in pure non-commutative QED  
is modified in the non-commutative directions 
by the factor $1 + \Lambda_{QCD}^4 \theta^2$,
where $\theta$ is the non-commutative parameter.
If charged flavours are included,  
there is an upper bound
on the momentum of a photon propagating in the non-commutative directions,  
beyond which it is unstable against production of charged pairs. 
We also discuss a particular $\theta\rightarrow\infty$ limit 
of pure non-commutative $QED$ which is T-dual to
a more conventional $N\rightarrow\infty$ limit with $m/N$ fixed.
In the non-commutative description, 
this limit gives rise to an exotic theory of open strings.  
 
\setcounter{page}{0}  
\newpage 
\setcounter{footnote}{0}


\section{Introduction}

In this paper,  we study strong coupling aspects of $U(1)$ gauge theory
on the non-commutative plane.  We emphasize that our intent is not
to examine possible experimental signatures of non-commutativity. 
It is known \cite{mpr,bd} that a non-commutative version of 
QED is inconsistent with experiment unless there is a cutoff 
which is small compared to the non-commutativity scale $1/\sqrt{\theta}$.
In the absence of a cutoff,  there are large lorentz violating operators
generated at one loop,  even for arbitrarily small
$|\theta|$, due to UV/IR mixing.  
It is also impossible to reproduce the observed
infrared stability of QED.  
The phenomena which we shall study arise in the absence of a cutoff, 
and are therefore primarily of theoretical interest.    
In this case the pure non-commutative $U(1)$ gauge theory is
asymptotically free \cite{MRS} and the effective coupling grows
at large distance, leading to interesting strong coupling effects. 

One tool which we will use to study such effects
is T-duality, or Morita equivalence.
On a non-commutative two-torus,  which we take to be square for simplicity, 
there is an  
$SL(2,Z)$ duality group relating $U(N)$ theories of different rank $N$, 
magnetic flux $m$, torus radius $R$,  and non-commutativity parameter 
$\theta$.
This duality is inherited from 
T-duality in string theory \cite{CDS,SW}, however  
it can be understood entirely in the context of 
Yang-Mills theory \cite{Sch, PS,B,AMNS,S,gt}.  

We shall begin by considering pure QED (no charged matter) 
on ${\mathcal R}^4$ with two non-commuting spatial directions;
$[x^1,x^2] = i\theta^{12} \equiv \theta$.
The gauge invariant photon creation operators in this theory
are open Wilson lines \cite{IS,DR,GHI,SD},  with seperation between
the endpoints given by $L^i = \theta^{ij}p_j$, where $p$ is the momentum.
The dispersion relation for photons at large momentum depends crucially
on whether or not these open Wilson lines have zero tension.
Pure non-commutative QED (NC-QED) 
shares much in common with conventional
QCD,  such as gauge field self interactions and asymptotic 
freedom \cite{MRS}. We shall argue that this similiarity also includes a
non-vanishing string tension.  The relation between NC-QED and QCD can be 
understood precisely by making use of T-duality.  
However, unlike asymptotic freedom,  the existence of a 
non-vanishing string tension in NC-QED is not manifest from this duality, 
since the dual description is a zero area limit of QCD on 
a torus with fixed $\Lambda_{\rm QCD}$.  Naively one would
expect the string tenstion to vanish and the electric flux to condense
below a critical radius.  

To make use of T-duality,  we shall view QED on the non-commutative
plane as the infinite area limit of QED on 
$T^2_{\theta} \times {\mathcal R}^2$,  where $T^2_{\theta}$ 
is a non-commutative two-torus.  This limit is T-dual 
to an $N \rightarrow \infty$ limit of $U(N)$ gauge theory
on a commutative $T^2 \times {\mathcal R}^2$.
There is a non-zero magnetic flux on the
dual $T^2$,  whose radius $R^{\prime}$ scales 
like $1/\sqrt{N}$ with $\Lambda_{QCD}$ fixed.  In the absence of  
magnetic flux, such a limit would lead to a truncation to the zero 
modes on the torus, i.e. dimensional reduction.
However because of the non-zero flux,  there are twisted 
boundary conditions
giving rise to fractional momentum modes which do not decouple
in the large N - small area limit. 
As will be clear later, the $U(1)$ part of the dual $U(N)$ theory 
plays a trivial role, 
describing the physics of a free decoupled sector of pure NC-QED.
The interesting dynamics resides in the $SU(N)$ component of the 
dual theory. 
 
In the $N\rightarrow \infty$ limit under consideration,  
the torus becomes small compared to $1/\Lambda_{QCD}$.
Nevertheless, one still encounters
strong coupling phenomena.  Strong coupling physics 
enters, for instance, into the computation of the energy of a 
sufficiently large electric flux on the torus.  This is somewhat 
counter-intuitive. If $N$ were fixed, the
energy of any electric flux on the torus could be computed 
perturbatively for a sufficiently small radius.  Perturbatively, this 
energy is zero when the third (commutative) spatial direction 
is non-compact.  One might naively expect electric flux 
condensation on a sufficiently small torus,  
via spontaneous breaking of the $Z_N \times Z_N$ 
group of large gauge transformations. 

The loophole is that, although we take a zero radius limit,  we 
also simultaneously take 
$N\rightarrow\infty$,  in which case
one can not neccessarily compute
the electric flux energy perturbatively.
At fixed $N$ there is a limit
to how large the electric flux can be,  
being defined only modulo $N$.
In the large $N$-small radius limit,
we may consider increasingly large electric fluxes whose energies 
are not perturbatively computable.  Furthermore,  the usual argument for 
$Z_N$ symmetry breaking at small radius 
does not apply,  due in part to the presence of a 
non-zero magnetic flux.   
We conjecture that in the absence of a $Z_N \times Z_N$ symmetry breaking 
transition,  
there is no critical
radius below which which the tension of QCD strings wrapping 
the torus vanishes.
In NC-QED,  this implies that open Wilson lines have a non-zero
tension.  The photons of NC-QED, which are created by 
such open Wilson lines,  are dual to 
QCD strings wrapping the torus 
$\sim \sqrt{N}$ times.  
Such strings
have fixed length and energy in the 
$N\rightarrow \infty$ limit with $R^{\prime} \sim 1/\sqrt{N}$. 

Since the length of open Wilson lines is proportional to their 
transverse momentum, a non-zero string tension leads to a dramatically 
modified dispersion relation for photons 
with large momentum in the non-commutative plane. At short 
wavelengths, we shall find that the speed of light in the 
non-commutative plane is modified by the factor
$1 + \Lambda^4|\theta|^2$,  where the dynamically generated scale 
$\Lambda$ is proportional to $\Lambda_{QCD}$ in the dual description.  
A similar phenomenon occurs at the {\it classical} level in 
non-commutative QED,  in which case the speed of light depends on the 
direction when a background magnetic field is turned on \cite{gjpp, cai}.

When charged matter is included in NC-QED, there is
an upper bound on the momentum of a photon propagating in a 
non-commutative direction.  At sufficiently large momentum, at
which the propagation speed is ``superluminal'', the photon is unstable 
and decays by $e_+ e_-$ pair production.
This decay is analagous to the instability of high energy photons in a 
strong magnetic field in convential QED, although the mechanism is quite 
different.   For sufficently large momentum,  the open Wilson line associated
with a photon becomes long enough that it is energetically favorable
to break off segments of Wilson line bounded
by electron-positron pairs,  the length of which is not constrained to be 
proportional to the transverse momentum.  
A similiar phenomenon would occur in
NC-QCD.  Glueballs in NC-QCD are also
open Wilson lines with length proportional to the momentum in
the non-commutative plane.  
Above a critical momentum,  they have an added instability against 
decay to mesons.  

While pure QED on the non-commutative plane
can be  obtained as an $N \rightarrow \infty$ limit of a conventional 
gauge theory,  this is not in general a planar limit.  It differs from the 
usual 't Hooft large $N$ limit because of the non-zero magnetic 
field and the $1/N$ scaling of the area of the torus area. 
In this case, finite size effects ruin
the usual large $N$ counting. The exceptions are the beta 
function and other quantities determined by UV singularities, 
which are independent
of the compactification and the magnetic flux,  and are therefore 
obtained entirely from planar graphs at $N\rightarrow\infty$.
  
In \cite{MRS} it was argued that
stringy behaviour of a non-commutative
field theory arises as $\theta \rightarrow \infty$,  
since non-planar graphs vanish in this limit.  
For a {\it gauge} theory, the physical content
of such a theory is not obvious.  
In the pure non-commutative gauge theory,  there are no 
local gauge invariant operators.  
Furthermore the open Wilson lines,  which are quasi-local operators,
have seperations satisfying 
$L^i = \theta^{ij}p_j$,  and are therefore ill defined in the 
$\theta\rightarrow\infty$ limit. 
Since only zero momentum operators survive
as $\theta\rightarrow\infty$,
space-time is in some sense absent from the string theory which 
arises in this limit.

There is however a modified 
large $\theta$  limit in which there exist well behaved 
gauge invariant operators with non-zero momentum.  
On a non-commutative torus,  open Wilson lines are characterized 
not just by their momentum,
but also by a ``winding number'' or electric flux.
In the limit $R\rightarrow\infty$, where $R$ is the radius of the torus, 
electric flux contributes an 
infinite amount to the length of the open Wilson line,  just 
as momentum does in the infinite $\theta$ limit.
The sum of these two infinite contributions may 
be finite provided $\theta$ and the area are
sent to infinity simultaneously.  In such a limit, one obtains 
an exotic open string theory.  The string coupling
of this theory is given by $l_s^2/\theta$,  where the string scale 
$l_s = \theta/R$ and is held fixed in the large 
$\theta$ limit.   

This open string theory is T-dual
to a conventional large $N$ limit of $U(N)$ gauge theory on a torus
of fixed size and fixed non-zero $m/N$, where $m$ is the magnetic flux.    
In some instances,  such as pure two dimensional NC-QED 
or strongly
coupled four dimensional ${\cal N} =4$ NC-QED,  it is possible to compute
correlation functions of open strings exactly to leading order
in the string coupling by using the dual $U(\infty)$ description.

The organization of this paper is as follows.  In 
Section II we give a brief review of gauge theory on a non-commutative
space and the T-duality symmetries of gauge theory on a non-commutative
torus. In Section III we construct the $U(\infty)$ dual
of pure QED on the non-commutative plane. In section IV we use this dual 
description
to argue that  open Wilson lines have non-zero tension, and compute
the photon dispersion relation. We then argue 
there is an upper bound on the photon momentum when charged matter is 
included. 
Section V contains a discussion of the difference between the 
$N\rightarrow\infty$ limit dual to NC-QED, and the usual 't Hooft 
$N\rightarrow\infty$ limit. 
In section VI we construct an open string theory from a 
$\theta \rightarrow\infty$
limit of NC-QED.
We then present some exact
open string correlators in the free string limit.     

\section{Review of non-commutative gauge theories \\ and Morita equivalence}
\label{review}

\subsection{Non-commutative gauge theory}

Field theory on a noncommutative space can be constructed
by replacing products of functions with
Moyal $*$-products.  The $*$-product is defined as follows;
\bea
f(x)*g(x) = e^{i \frac{\theta^{\mu\nu}}{2}
\frac{\partial}{\partial x^{\mu}}\frac{\partial}{\partial y^{\nu}}}
        f(x)g(y)|_{y\rightarrow x}
\label{stardef}
\eea
The action of the pure noncommutative $U(p)$ Yang-Mills theory is
\bea
S=\frac{1}{g^2} \int d^Dx 
Tr({F}_{\mu\nu}(x) * {F}^{\mu\nu}(x))
\eea
where 
\bea
{F}_{\mu\nu}(x) = \partial_{\mu}{ A}_{\nu} - 
\partial_{\nu}{A}_{\mu} - 
i ({A}_{\mu} * {A}_{\nu} - 
        {A}_{\nu}* {A}_{\mu})
\eea
In much of what follows,  we shall take $\theta_{12} = - \theta_{21} =\theta$
to be non-zero with all other components vanishing.  
When the $1-2$ plane is compactified on a torus of 
radius $R$,
the non-commutative parameter becomes a periodic variable
$\theta \sim \theta + 4\pi R^2$,  and it will be convenient to define
the dimensionless quantity $\Theta \equiv \theta/(4\pi R^2)$.  

The pure non-commutative gauge theory has no local gauge invariant operators. 
Instead, there are quasi-local gauge invariant operators
constructed from open Wilson lines with
possible insertions of local gauge covariant operators \cite{IS,DR,GHI,SD}. 
The open Wilson lines are gauge invariant provided the seperation
between the endpoints is $L^i = \theta^{ij}p_j$, where $p$ is the
momentum.  

As an example, a  
straight open Wilson line in the $2$ direction of the non- commutative
$1-2$ plane is given by 
\bea
\int dx^1dx^2 P^{*} \left(e^{i\int_C A_2}\right) * e^{ip_1x^1}
\eea 
where $C$ is a path between the points $(x^1,x^2,x^3)$ and
$(x^1,x^2 + \theta^{12}p_1,x^3)$,  and $P^*$ indicates that the exponential
is path ordered with respect to the non-commutative $*$-product. 
An example of a gauge invariant operator 
with a non-trivial $\theta\rightarrow 0$ limit
is
\bea
\int d^3x P^{*} \left(F_{\mu\nu}e^{i\int_C A_2}\right) * 
e^{i{\bf p}\cdot{\bf x}}
\label{insert}
\eea
where the local operator $F_{\mu\nu}$ is inserted anywhere
within the  open Wilson line.

On a torus the open Wilson line may also have an 
integer winding number 
(electric flux) $e_i$ \cite{gt}. Gauge invariance again requires
the endpoints to be seperated by 
$L^i  = \theta^{ij}p_j$,  however the minimum length with an
electric flux $e_i$ is $\sqrt{{\cal L}_i{\cal L}_i}$ where 
${\cal L}_i = \theta^{ij}p_j + 2\pi R e^i$ and $R$ is the 
radius of the (square) torus.
Non-commutative electric fluxes will play a role in
our discussion of the $\theta\rightarrow \infty$ limit.   

\subsection{T-duality}
\label{morita}

$U(N)$ gauge theory on a non-commutative $T^2$ exhibits an $SL(2,Z)$ 
Morita equivalence which is 
inherited from string theory T-duality (see \cite{CDS, SW}),  but 
which can be demonstrated explicitly without recourse to 
string theory \cite{Sch,PS,B,AMNS,S,gt}.  
This is not a duality in the traditional sense, as it exists at the
classical level.  The duality acts as follows: 
\bea
\begin{pmatrix} N \\ m  \end{pmatrix} &\rightarrow&
\begin{pmatrix} a & b  \\ c & d \end{pmatrix}
\begin{pmatrix} N \\ m \end{pmatrix} \label{a}\\
\Theta &\rightarrow& 
\frac{c + d\Theta}{a + b\Theta} \label{b}\\
R^2 &\rightarrow& R^2(a+ b \Theta)^2 \label{c}\\
g^2 &\rightarrow& g^2(a + b\Theta) \label{d}\\
\Phi &\rightarrow& (a+b\Theta)^2 \Phi - b(a + b\Theta)
\label{e}
\eea
where $\Theta \equiv \theta/ (4\pi R^2)$,
$m$ is the magnetic flux, 
$g$ is the gauge coupling, and $R$ is the radius of the torus.
The action is
\bea
S = \frac{1}{g^2}\int d^Dx{\rm tr} (F_{\mu\nu} - 2\pi \Phi_{\mu\nu} I)^2.
\eea
where $\Phi_{\mu\nu}$ is a constant antisymmetric tensor background 
for which all components vanish except those with indices on the 
non-commutative torus $\Phi_{12} = \Phi_{21} \equiv \Phi$.

\section{The non-commutative plane from an $N \rightarrow \infty$ limit}

We will study $U(1)$ gauge theory on the non-commutative plane by
taking an infinite volume limit of the theory on a non-commutative
torus of radius $R$.  The first Chern class and 
background term $\Phi$ shall be set to zero.  
We consider a rational non-commutative
parameter, 
$\Theta = -c/N$, where $c$ and
$N$ are integers with greatest common divisor one. 
When taking the infinite volume
limit,  $\theta = 4\pi R^2\Theta$ will be held fixed.
There is no loss of generality in considering rational $\Theta$,  
since one can still obtain any $\theta = 2\pi R^2\Theta$ in the  
infinite $R$ limit.  

For $\Theta = -c/N$,  there is a dual description in terms of
a $U(N)$ gauge theory on a commutative torus with non-zero magnetic flux.
The magnetic flux $m$ is a solution of the equation $aN - cm = 1$ 
for integer $a$. 
The relations (\ref{a}) - (\ref{e}) then become
\bea
&&\begin{pmatrix} 1 \\ 0  \end{pmatrix} \rightarrow
\begin{pmatrix} a & m  \\ c & N \end{pmatrix}
\begin{pmatrix} 1 \\ 0 \end{pmatrix} = 
\begin{pmatrix} N \\ m \end{pmatrix} \label{aa}\\
&&\Theta = -\frac{c}{N} \rightarrow 
\Theta^{\prime}= \frac{c + N\Theta}{a + m\Theta} = 0 \label{bb}\\
&&R^2 \rightarrow {R^{\prime}}^2 = R^2(a- m \frac{c}{N})^2 = 
\frac{R^2}{N^2} \label{cc}\\
&&g^2 \rightarrow {g^{\prime}}^2 = g^2(a - m\frac{c}{N}) = 
\frac{g^2}{N} \label{dd}\\
&&\Phi =0 \rightarrow \Phi^{\prime} = - m(a - m\frac{c}{N}) = -\frac{m}{N}
\label{ee}
\eea
The equation $aN - cm = 1$ only fixes $m$ modulo $N$.  
For a $U(N)$ bundle, the  first Chern class $m$ is 
equal, modulo $N$, to the 't Hooft magnetic flux in the 
$SU(N)/Z_N$ sector of the theory,  which is defined only modulo $N$ 
(see for instance \cite{GR}).  In the $U(N)$ description, the background 
term $\Phi^{\prime}$ 
is equal to the first Chern class, so that 
$\int_{T^2} tr(F^{\prime} - \Phi^{\prime} I) = 0$. 
There is a symmetry which simultaneously shifts $m$ by $N$ and 
the $\Phi$ by $1$. 
This symmetry relates all the solutions of $aN-cm =1$ 
for a given $c$ and $N$.  
Note that the action is the same as that of
a $U(1) \times SU(N)/Z_N$ gauge theory in a sector 
with vanishing magnetic flux in the $U(1)$ component
and 't Hooft flux $m$ in the $SU(N)/Z_N$ component.
The dynamics in the $U(1)$ component is
trivial.  This sector is dual to a free decoupled sector of pure NC-QED 
containing photons with momenta $Nk_i/R$ in the non-commutative plane, where
$k_i$ are integers.   
The interesting dynamics resides is in the $SU(N)/Z_N$ component.

In order to take the $R\rightarrow\infty$  limit with 
$\theta = \Theta 4\pi R^2$ fixed, one must 
take $R^{\prime}N \rightarrow \infty$ keeping $c{R^{\prime}}^2N$ fixed.   
One way to do this is to 
take the $N\rightarrow \infty$ limit with  $c=1$, $m =N-1$ and 
$4\pi {R^{\prime}}^2 = \theta/N$. 
Note that ${g^{\prime}}^2N$ is equal to the NC-QED coupling $g^2$ and 
should therefore be held fixed 
in the large $N$ limit.  If not for the
$1/\sqrt{N}$ scaling of the radius $R^{\prime}$,  this would be a conventional
't Hooft large $N$ limit.  $\Lambda_{\rm QCD}$ will be
held fixed as $N\rightarrow\infty$.  This is neccessary so that 
the dynamically generated scale $\Lambda$ in the non-commutative
$U(1)$ description is also held fixed in the large $R$ limit.
At the quantum level, the relation $g^2 = {g^{\prime}}^2N$ is
consistent if the couplings are evaluated at the same scale.
To one loop, the running coupling of the 
non-commutative $U(1)$ theory is \cite{MRS}
\bea
\frac{1}{g^2(\mu)} = \frac{1}{16\pi^2}\frac{11}{3}{\rm ln}\frac{\mu}{\Lambda}
\eea
while that of the 't Hooft coupling $\lambda = {g^{\prime}}^2N$ in
the $U(N)$ dual is
\bea
\frac{1}{\lambda(\mu)} = \frac{1}{16\pi^2}\frac{11}{3}{\rm ln}
        \frac{\mu}{\Lambda_{\rm QCD}}.
\eea
Thus $\Lambda = \Lambda_{\rm QCD}$.  
It is known \cite{bigsuss} that the beta function of pure non-compact 
NC-QED is the same as that of the 
$N\rightarrow\infty$ limit of pure $SU(N)$  gauge theory to all orders in
perturbation theory\footnote{Due to UV/IR mixing,  short distance 
singularities
in a non-commutative theory may depend on the compactification.
In the dual $SU(N)$ description, this can be seen explicitly in  
$1/N$ corrections to the running of the 't Hooft coupling.}.
Like conventional QCD,  Feynman graphs in a 
non-commutative theory can be written in a double line notation and 
organized into a genus expansion \cite{doubline}.    
The relation between the beta function of pure uncompactified 
NC-QED and that of the $N\rightarrow \infty$ commutative theory 
follows from the fact that UV divergences in the uncompactified 
non-commutative theory are due to planar graphs\footnote{This relation
between the beta functions,  like the duality, 
is invalid when charged matter is included in NC-QED.}.   

Although the radius in the $U(N)$ dual of NC-QED is scaled to zero 
in the large $N$ limit, 
this scaling does not amount to a dimensional reduction.  The reason
is that the gauge fields in the $U(N)$ description $A^{\prime}$ 
satisfy twisted boundary conditions which permit fractional momenta.  
It will be convenient to define the field
\bea
{\tilde A}^{\prime}_{\mu} \equiv A^{\prime}_{\mu} - 
\delta_{\mu 1}\frac{m x^2}{N 2\pi {R^{\prime}}^2} I
\eea
We will choose boundary conditions such that
${\rm tr}{\tilde A}^{\prime}_{\mu}$ 
is periodic,  and
\bea
{\tilde A}^{\prime}(x^1+2\pi R,x^2) = 
P^{-m} {\tilde A}^{\prime}(x^1,x^2) P^{m} \\ 
{\tilde A}^{\prime}(x^1, x^2 + 2\pi R) = 
Q {\tilde A}^{\prime}(x^1,x^2) Q^{-1}
\label{bcs}
\eea
where  
\bea
Q =
\begin{pmatrix} 1 & & & \\
& e^{\frac{2\pi i}{N}} & & \\
& & \ddots & \\
& & & e^{\frac{2\pi i (N-1)}{N}} \end{pmatrix}
\label{defq}
\eea
and
\bea
P = 
\begin{pmatrix}
 0 & 1 & & \\
 & 0  & 1 & \\
 &   &   & \ddots \\
 1 &   &   &
\end{pmatrix}  
\label{defp}
\eea
$Q$ and $P$ satisfy 
\begin{equation}
PQ = QP e^{2\pi i/N}.
\end{equation}
These boundary conditions correspond to a sector with 't Hooft magnetic
flux (and first Chern class) $m$.
Since $(P^m)^N = Q^N = 1$,  the gauge field ${\tilde A}^{\prime}$
is strictly periodic on an $N^2$ cover of the torus,  with periods
$2\pi NR$.  Because the greatest common divisor of $N$ and $m$ is $1$ in
our case, there is no smaller multiple cover on which the gauge fields 
are periodic.  Note that the $U(1)$ ($\sim$ identity) component of 
${\tilde A}^{\prime}$ 
is periodic on the torus so that, in the large $N$ limit which we consider,
only the zero modes of the $U(1)$ sector survive. 
In the $SU(N)/Z_N$ sector however,  the gauge field can not be written
as a sum of periodic fluctuations and a non-periodic background.
There are modes of finite (fractional) momenta which do not decouple
in the large $N$/small radius limit.

For completeness, the explicit map \cite{AMNS,S,gt} between the 
gauge field $A$ of non-commutative QED gauge 
and the $U(N)$ gauge field $A^{\prime}$ is as follows.
A general solution of the boundary conditions (\ref{bcs}) is
\bea
{\tilde A}^{\prime} =  
\sum_{\vec r} a^{\mu}_{\vec r} Q^{-c r_1}
P^{r_2} \exp \left(-i\pi \frac{c}{N}r_1r_2 \right)
\exp \left( - i \frac{ {\vec r} \cdot {\vec x} } {NR^{\prime}}\right).
\label{genform}
\eea
where
the independent variables $a^{\mu}_{\vec r}$ are the fourier modes
of the non-commutative gauge field,
\bea
A = \sum_{\vec r} a^{\mu}_{\vec r} 
\exp(-i \frac{ {\vec r} \cdot {\vec x} }{R}).
\eea

Despite the twisted boundary conditions, the gauge invariant 
operators of the $U(N)$ theory obey 
periodic boundary conditions.  These operators can in general 
be split into $U(1)$ and $SU(N)$ parts.  These parts
are seperately periodic for gauge invariant local 
operators and Wilson loops with trivial homology.  
With the exception of modes with zero momentum on the torus,  
such opertators create 
states with infinite energy in the $N \rightarrow \infty$ limit
with $R^{\prime} \sim 1/\sqrt{N}$, and may be integrated out.  
On the other hand,  the Wilson loops with non-trivial homology
factorize into $U(1)$ and $SU(N)/Z_N$ parts which have fractional 
momenta.  These operators will be of particular interest, as they
correspond to open Wilson lines in the 
non-commutative description \cite{S,gt,AMNS}.  
Due to the fractional momenta of the seperate components,
these operators do not decouple in the large $N$ - small radius limit.

We will now make the above  discussion more precise. 
In the $U(1)$ sector of the $U(N)$ theory,  
the gauge field can be written as a sum of a 
non-periodic background term 
(accounting for the non-zero first Chern class) and a periodic piece 
which is integrated out, with the exception of the zero modes on the 
torus. Consider a Wilson loop in the fundamental representation 
of $U(N)$ which wraps the torus  $e_2$ times in the $2$ direction.
The holonomy is given by
\bea
W_{e_2}(x^1) = {\rm tr}\left[ P e^{i\oint A^{\prime}_2} \Omega_2^{e_2} \right]
\label{thop}
\eea
where $\Omega_2$ is a $U(N)$ transition function associated
with the twisted boundary conditions in the $2$ direction, and
is required for gauge invariance \cite{VB}.  
For the boundary conditions we have chosen,
\bea
\Omega_2^{e_2} = P^{-e_2}e^{-i\frac{me_2x^1}{NR}}.
\eea 
The operator (\ref{thop}) is periodic in $x^1$, and one can define
Wilson loops with transverse momentum $n_1/R^{\prime}$;
\bea
{\cal W}_{e_2, n_1} \equiv \int_0^{2\pi R^{\prime}}dx^1 W_{e_2}(x^1)
        e^{i\frac{n_1x^1}{R^{\prime}}}
\eea
where $n_1$ is an integer.
Splitting this into $SU(N)$ and $U(1)$ components gives
\bea
{\cal W}_{e_2, n_1} =
e^{iC_2(x^0,x^3)}\int_0^{2\pi R^{\prime}}dx^1 
e^{i\frac{(n_1 + \frac{m}{N}e_2) x^1}{R^{\prime}}} 
{\tilde {\cal W}}_{e_2}(x^1)
\eea
where $\tilde W$ is the $SU(N)$ component of the holonomy,  
involving $SU(N)$ valued transition functions, and
$C_{\mu}(x^0,x^3)$ is the zero mode of the $U(1)$ component of
the gauge field;
\bea
C_{\mu}(x^0, x^3) = 
\frac{1}{N}\int_{T^2}{\rm tr} {\tilde A}^{\prime}_{\mu}
\eea
The $U(1)$ field ${\rm tr} {\tilde A}^{\prime}$ is periodic, so 
except for the zero modes, the $U(1)$ sector decouples in the large N limit
with $R^{\prime} \sim 1/\sqrt{N}$.
On the other hand, the $SU(N)$ component of the holonomy $\tilde W(x^1)$ is 
periodic only over an $N$-fold cover of the 
torus, 
\bea
{\tilde W}_{e_2}(x^1 + 2\pi R^{\prime}) = {\tilde W}_{e_2}(x^1)
e^{2\pi i \frac{m}{N}e_2}
\eea
We can therefore define (fractional) fourier modes
\bea
{\tilde {\cal W}}_{e_2, k_1} 
\equiv \frac{1}{N} \int_0^{2\pi R^{\prime} N} dx^1 
        {\tilde W}(x^1) e^{i\frac{k_1x^1}{R^{\prime}}}
\eea
with $k_1$ an integer multiple of $1/N$. Then
\bea
{\cal W}_{n_1,e_2} = 
e^{-iC_2}{\tilde {\cal W}}_{e_2, n_1 - \frac{m}{N}e_2} 
\eea 
The fractional momentum 
$(n_1^{\prime} - \frac{m}{N}e_2^{\prime})/R^{\prime}$ 
may be finite in the large $N$ limit with $R^{\prime}\sim 1/\sqrt{N}$,  
so the states associated with Wilson loops of non-trivial homology 
can not neccessarily be integrated out. 
For $m = N-1$,  one can get finite momentum by
choosing $n_1= e_2 \sim \sqrt{N}$.

\section{The non-commutative photon}

\subsection{The dual of the non-commutative photon}

We now consider photon propagation in three spatial dimensions,
with non- commutativity in the $12$ plane.  On the commutative 
plane,  the photon is created by a local gauge invariant operator 
such as ${F}_{\mu\nu}$.  On the non-commutative plane the natural 
generalization is the open Wilson line with an insertion of
${F}_{\mu\nu}$, as in (\ref{insert}).
    
The explicit map between open Wilson lines in non-commutative QED at
rational $\Theta$ and operators of the dual commutative $U(N)$ theory 
is discussed in
\cite{gt, S, AMNS},  and we review it briefly below. 
For simplicity, we restrict ourselves
to open Wilson lines in the $2$ direction,  which are dual to 
$U(N)$ Wilson loops in the fundamental representation wrapping 
in the $2$ direction. The open Wilson lines are 
characterized
by integers $(n_1,e_2)$ labelling the transverse momentum and 
winding (or electric flux) respectively.  Similiarly the holonomies of
the commutative theory are characterized by a transverse momentum 
and winding number $(n_1^{\prime}, e_2^{\prime})$.  
Note that $n_1^{\prime}$ is the   
momentum of the full $U(N)$ 
holonomy,  which is integer rather than fractional.  
The map between the non-commutative open Wilson lines and the 
commutative holonomies is given by
\bea
\begin{pmatrix} e_2 \\ n_1  \end{pmatrix} =
\begin{pmatrix} a & -c \\ -m & N \end{pmatrix}
\begin{pmatrix} e_2^{\prime} \\ n_1^{\prime} \end{pmatrix}=
\begin{pmatrix} 1 & -1 \\ 1-N & N \end{pmatrix} 
\begin{pmatrix} e_2^{\prime} \\ n_1^{\prime} \end{pmatrix}
\label{map}
\eea
This map is valid with or without local operator insertions.

Since we are considering the large volume limit of the non-commutative
theory,  we will take $e_2 =0$.  Via (\ref{map}),  this implies 
$e_2^{\prime} = n_1^{\prime} = n_1$.  We wish to keep the   
momentum $p_1= n_1/R$ fixed in the large $N$ limit with $R \sim \sqrt{N}$,
therefore  
$n_1 = n_1^{\prime} = e_2^{\prime}\sim \sqrt{N}$.
The wrapped Wilson loop in the $U(N)$ description  has length
$2\pi R^{\prime}e_2^{\prime}$ which is fixed in the large $N$ limit,
and equal to that of the open Wilson line in the non-commutative
description.  The $SU(N)$ component of the Wilson loop carries a fractional
momentum $p^{\prime}_{SU(N)}= 
(n_1^{\prime} - \frac{m}{N}e_2^{\prime})/R^{\prime}$.  This momentum is equal 
to that of the dual open Wilson line, $p_1 = n_1/R$. 
To summarize,  the relation between non-commutative QED and 
a large $N$ limit of $U(N)$ gauge theory is given by 

\vspace{30pt}

\begin{tabular}{|l|l|}
\hline
NC-QED on $T^2_{\theta} \times {\mathcal R}^2$  & 
                $U(N)$ YM on $T^2 \times{\mathcal R}^2$  \\
\hline
magnetic flux $ 0$ & $m=N-1$ \\
\hline
radius $R = \sqrt{\theta N}$, $\theta$ fixed 
& $R^{\prime} = \sqrt{\theta/N}$ \\
\hline
coupling $g^2$ fixed & ${g^{\prime}}^2 N = g^2$ \\
\hline
$p_1 = \frac{n_1}{R}$ fixed,  & 
${p_1}^{\prime}_{SU(N)/Z_N} =  p_1 = 
\frac{n_1^{\prime} - \frac{m}{N}e_2}{R^{\prime}}$ \\ 
$e_2 =0$ & $n_1^{\prime} = e_2^{\prime} = n_1 \sim\sqrt{N}$ \\   
\hline
\end{tabular}
\vspace{10pt}

Table 1. 

\subsection{Non-vanishing string tension}

The energy of a photon propagating on the 
non-commutative plane should depend on the momentum, $\theta$
(and perhaps the polarization). 
For simplicity,  we take the momentum to be in 
the $1$ direction.  In the
$U(N)$ description, we must compute the energy 
of a state created by a fundamental Wilson loop of non-trivial homology, 
with transverse momentum $n_1^{\prime}$ and winding number $e_2^{\prime}$ 
satisfying $n_1^{\prime} = e_2^{\prime} \sim \sqrt{N}$. 
The length of the Wilson loop, $2\pi e_2^{\prime} R^{\prime}$ is fixed 
in the large $N$ limit,  and is equal to the length of the open Wilson line
in the non-commutative description.  In order for this limit to be smooth,
the QCD string tension should be fixed as $N\rightarrow \infty$ with 
$R^{\prime} \sim 1/\sqrt{N}$.  The dispersion relation depends in 
large part on whether this tension vanishes or not.  Naively,  one
would expect the answer to depend on the ratio the radius of the 
torus to the QCD length scale, $R^{\prime}\Lambda_{\rm QCD}$.  
This ratio goes to zero in the limit which we consider, since
$\Lambda_{\rm QCD}$ is held fixed.

One would naively expect the electric flux on the torus to condense
when the radius $R^{\prime}$ 
becomes sufficiently small compared to
$1/\Lambda_{QCD}$.  
Electric flux condensation corresponds to spontaneous breaking of
the  $Z_N \times Z_N$ 
symmetry of large gauge transformations on the torus,  since 
states with definite electric flux are eigenstates of these transformations.
In a more conventional scenario in which $N$ is fixed
and a single compact dimension is shrunk,  there is
a $Z_N$ symmetry breaking transition analagous to a finite
temperature deconfinement transition \cite{YS}.  
In our scenario however, the usual arguments for  
$Z_N$ symmetry breaking at small radius do not apply. 
In the absence of such a transition,  one expects the string tension
to remain non-zero even at small radius. 

To see why electric flux condensation should not be
expected for $N\rightarrow \infty$ with $R^{\prime} \sim 1/\sqrt{N}$, 
let us first recall the argument of \cite{YS},  which appies to an $SU(N)/Z_N$
theory in the absence
of magnetic flux,  and which involves shrinking the 
compactified Euclidean time direction. Note that when the compact 
direction is time,  the interpretation of $Z_N$ symmetry breaking 
is different, i.e.
finite temperature deconfinement.
When the compactified directions is
spatial,  $Z_N$ symmetry breaking implies condensation 
of electric flux in the compactified direction.
At high temperature, the modes on the $S^1$ may be taken 
to be constant. The action can then be written
\bea
S = \frac{1}{g^2}\int d^3x{\rm tr}
        [\frac{1}{\beta}D_ihD_ih^{\dagger} + \beta F_{ij}^2]
\label{pint}
\eea
where $\beta = 2\pi R$ is the radius of the $S^1$, and
$h$ is the $SU(N)$ valued holonomy in the time direction,
$h= P \exp(i\oint A_0)$. When restricted to the zero mode,
$h= \exp(i\beta A_0)$.
The $Z_N$ center symmetry acts as $h\rightarrow zh$ where 
$z \epsilon Z_N$.  A gauge invariant order parameter is
$\rho \equiv {\rm tr} h$.  
At small $\beta$,  the path integral is dominated by configurations
with $h^{\dagger}D_ih = 0$.  This equation is satisfied for a generic  
$A_i$ only if the holonomy $h$ is a constant element in
the center of $SU(N)$. 
For sufficiently small $\beta$ the effective 
potential for $\rho$ has minima at traces of elements of $Z_N$,  implying
that the $Z_N$ symmetry is spontaneously broken.

The loophole to the above arguments in the case
in which there is a non-zero magnetic flux, $m=N-1$, and two $S^1$'s are 
shrunk to zero with $R^{\prime}\sim 1/\sqrt{N}$, is
that one can not truncate to the zero modes in the compact
directions.  This truncation is impossible due to 
fractional momenta quantized in units of 
$1/N$. Note that even if the radius were taken to zero at fixed 
$N$, which would allow such a truncation,  it is not clear that 
$Z_N \times Z_N$ symmetry breaking should occur.  
When the {\it two} $S^1$'s are shrunk
simultaneously,  the reduced action is
\bea
S = \frac{1}{g^2}\int d^2x{\rm tr}
[D_ih_{\kappa}D_ih^{\dagger}_{\kappa} +\beta^2 F_{\alpha\beta}^2]
\eea  
where $\alpha,\beta = 0,3$ are the uncompactified directions and 
$\kappa = 1,2$ are the compactified directions.
Due to the diferrent scaling of the terms involving $h_i$, the
small $\beta$ limit does not give a path integral dominated
by configurations with vanishing
$h_i^{\dagger}D_ih_{\kappa}$.    

The $Z_N \times Z_N$ large gauge transformations on the commutative 
torus are dual to to discrete 
translations\cite{gt} on the non-commutative torus.  This can
be seen from the relation between their respective eigenvalues;
\bea
e^{2\pi i \frac{e_2^{\prime}}{N}} = e^{2\pi i \frac{n_1}{N}}
\eea
Thus electric flux condensation in the $U(N)$ description would
require spontaneous breaking
of translation invariance in pure NC-QED.
Although there are examples of of spontaneously broken 
translation invariance
in non-commutative {\it scalar} field theories \cite{gubs},
the above discussion suggests that this not the case in NC-QED.

The energy of a fixed electric flux on the torus can be
computed  perturbatively when the radius of the torus is scaled to 
zero for fixed $\Lambda_{QCD}$.  Perturbatively the energy of 
an electric flux is zero if the third spatial direction transverse to 
the 2-toris non-compact.  
However since the small radius limit is accompanied by a large $N$ limit, 
$R^{\prime}{\sim} 1/\sqrt{N}$,  in which case one may consider increasingly
large electric fluxes, which are defined modulo $N$. 
The energy of a sufficiently large electric flux is not 
computable by semiclassical means. 
We are interested in electric fluxes $e_i^{\prime}$ 
scaling like 
$\sqrt{N}$ (see table.1),  since these are dual to excitations in
NC-QED with finite momentum in the non-commutative plane.  In the 
$SU(N)/Z_N$ description,  these correspond to Wilson loops wrapping
the torus $\sim \sqrt{N}$ times.   
The contribution of the string tension $\sigma$ to the energy of such a state
is $E\sim \sigma e_i R^{\prime}$, which is finite
as $N\rightarrow \infty$ if string tension is fixed in this limit.  
In the absence of a $Z_N \times Z_N$
symmetry breaking transition,  it seems a reasonable conjecture that,
up to a numerical factor,  the string tension satisfies 
$\sigma = \Lambda_{\rm QCD}^2$, which is fixed in the large
$N$ limit. 

Note that for non-trivial physics to occur in the non-commutative
theory at finite momenta,  the relevent scale at which the coupling is
evaluated  in the dual $U(N)$ theory can not be set by 
$1/R^{\prime}$.  At this scale the coupling vanishes as $N\rightarrow \infty$.
For this purpose the large $N$ limit is essential.  
For the computation of the energy of
an electric flux,  the relevent scale is $1/(e_i^{\prime}R^{\prime})$
which is less than the QCD scale for sufficiently large electric fluxes

\subsection{The dispersion relation}

We seek the photon dispersion relation 
$E({\bf p}, \theta, \Lambda)$ where $E$ is the energy.
Let us first consider an open Wilson line with momentum $p_1$ in
the $1$ direction of
of the non-commutative torus, which is dual to a Wilson loop of the
same length wrapping the $2$ direction of the dual torus.  
If the length $\theta p_1$ of the open Wilson line 
is much greater than $1/\Lambda$,  then the length of the dual QCD string
is large compared to $1/\Lambda_{\rm QCD}$. 
In this regime one expects,  based on the previous arguments 
suggesting a non-zero
string tension,  that the dominant 
contribution to the energy is given by
$E \sim \Lambda^2p_1 \theta$,
assuming that $\Lambda^2 \theta p_1 >> p_1$. 

Let us also consider a photon propagating
in the $3$ direction transverse to the non-commutative plane.
Modes which carry no momentum in the non-commutative directions
are decoupled and free,  as can be easily seen from the definition of
the $*$-product (\ref{stardef}). 
The dispersion relation for such a photon
is $E= |p_3|$. Note that in the dual $U(N)$ description,  
this corresponds to an
excitation in the decoupled $U(1)$ sector.  
The energy spectrum in this sector is given by
\bea
E =\sqrt{\left(\frac{k_1}{R^{\prime}}\right)^2 + 
\left(\frac{k_2}{R^{\prime}}\right)^2 + p_3^2}
\eea
where $k_1$ and $k_2$ are integer momenta.
In the small $R^{\prime}$ limit,  the only $U(1)$ modes which
survive are those which have momentum only in the $3$ direction.

The dispersion relation should be consistent with
the above results and reduce to $p_{\mu}p^{\mu} = 0$ in 
the weak coupling limit
$\Lambda^2 \theta \rightarrow 0$.  It should also
be consistent with the violation of lorentz invariance by
the rank two tensor $\theta^{\mu\nu}$. This suggests the 
dispersion relation 
\bea
p^{\mu}p_{\mu} \sim \Lambda^4 
p_{\alpha} \theta^{\alpha\beta} \theta_{\beta\gamma} p^{\gamma},
\label{pole}
\eea
There may be subleading corrections to this which become important
in the weak coupling regime 
$p_{\alpha} \theta^{\alpha\beta} \theta_{\beta\gamma} p^{\gamma} << 
\frac{1}{\Lambda^2}$. In this regime, the dispersion relation should 
be calculable by perturbative means,  although we will not do so 
here. 

A subtle point is that the definition of the four momentum 
$p_{\mu}$ for 
$\theta_{0i} \ne 0$ requires some care,  since the usual Hamiltonian 
formulation is not applicable.  
We interpret the expression (\ref{pole}) 
as giving a pole in a propagator computable in a 
path integral formalism.   
For non-zero $\theta_{12} = -\theta_{21} =\theta$ with all other
components vanishing, the dispersion relation (\ref{pole}) 
is
\bea
p_0 = \sqrt{  (p_1^2 + p_2^2)(1 + \Lambda^4\theta^2) + p_3^2}.
\eea
In writing this expression,  we have absorbed possible numerical
factors into the definition of $\Lambda$.
In the strong coupling (or large photon momentum\footnote{It is difficult to 
probe the asymptotically free regime 
with gauge invariant operators in NC-QED, due 
to UV/IR mixing.  There is a large length scale associated with high 
momentum photons.}) regime in which the 
above expression is valid,  the speed of light
in the non-commutative plane is modified
by the factor $(1 + \Lambda^4\theta^2)$. 
Note that this factor is 
always greater than one, so that very high frequency light 
propagates ``superluminally'' in the non-commutative directions.

\subsection{Including matter}

We now consider the non-commutative $U(1)$ 
gauge theory with charged matter.  
At energies below the mass of the lightest charge matter,
the effective theory is the pure non-commutative $U(1)$ theory 
discussed above,  with a a dynamically generated scale $\Lambda$ and
a string tension $\sim \Lambda^2$.  
For sufficiently large momentum  
in the non-commutative plane,  we expect the photon to become unstable
and decay by production of charged pairs.  

This instability becomes kinematically
possible due to the modified dispersion relation (\ref{pole}).
The argument giving the modified dispersion relation
for the photon (\ref{pole}) is not applicable to charged 
matter,  since the length of open Wilson lines bounded by charged 
sources is arbitrary and unrelated to the transverse momentum.  
We shall assume that unlike the photon, the electron has
approximately the usual dispersion relation $p^2 = m_e^2$ 
at large momentum. 
Sufficiently long open Wilson 
lines, corresponding to high momentum photons,  
are then unstable against the 
production of electron-positron pairs via the process 
illustrated in figure 1.  This is analogous to pair 
production in QCD as quark anti-quark pairs are pulled apart.
Note that a similiar instability occurs in conventional 
commutative
QED when high energy photons propagate in a background magnetic field,
although the mechanism is quite different.

Given the dispersion relation (\ref{pole}), 
the stability bound on the momentum of a photon in non-commutative QED 
is given by
\bea
\Lambda^4 
p_{\alpha} \theta^{\alpha\beta} 
\theta_{\beta\gamma} p^{\gamma} < 4m^2
\label{QEDbnd}
\eea
where $m$ is the mass of the lightest charged particle and
$\Lambda$ the dynamically generated scale of the effective
theory at energies below $m$.

Analougous arguments also apply to the stability of 
glueballs in non-commutative QCD.  Like photons in NC-QED, 
Glueballs in NC-QCD are created by open Wilson lines,  
with length proportional to 
the momentum in the non-commutative plane. In addition the 
usual instabilities, there is an enhanced instability beyond a 
critical momentum satisfying 
\bea
\Lambda^4 
p_{\alpha} \theta^{\alpha\beta} 
\theta_{\beta\gamma} p^{\gamma} = m_{\pi}^2
\label{QCDbnd}
\eea
where $m_{\pi}$ is the mass of the lightest pion 
and $\Lambda^2$ is the QCD string
tension.  Note that unlike NC-QED,  the string tension remains 
non-zero even when the cutoff is small compared to the non-commutativity 
scale.  However such a cutoff excludes momenta 
where this instability could be observed.

\begin{figure}
\begin{center}
\epsfig{file=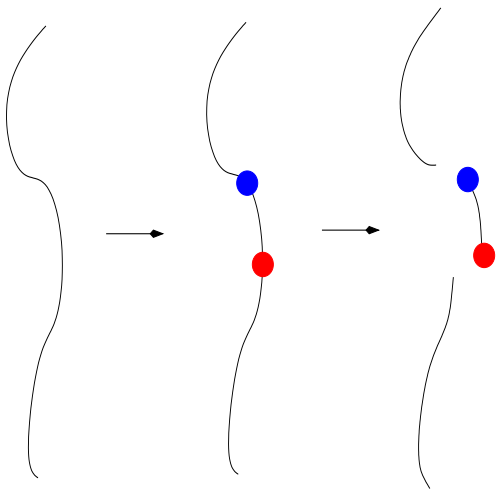, height=7cm,width=7cm}
\end{center}
\caption{At large momentum,  photons (open Wilson lines) in NC-QED 
with a single charged flavor are unstable. The process illustrated is 
$\gamma \rightarrow \gamma\gamma e_+ e_-$.  This instability also 
exists for large momentum glueballs in NC-QCD,  in which case the 
figure above corresponds to $G \rightarrow G G \pi$} 
\label{figure1}
\end{figure}

\section{'t Hooft large $N$ limit vs large N - small radius limit.}

The $N\rightarrow\infty$ limit of commutative $U(N)$ gauge theory
which is dual to the non-commutative 
$U(1)$ theory differs from the convenional 't Hooft limit in the 
$N$ dependent scaling of the compactification radius.  As 
a consequence of this scaling,  the $N\rightarrow\infty$ limit is 
not a planar limit for most observable quantities.  
Some exceptions are the beta
function and anomalous dimensions,  which are independent of 
the compactification in the commutative $U(N)$ theory. 
This is consistent
with what one expects for a $U(1)$ theory on the non-commutative
plane at finite $\theta$,  in which observables in general 
recieve contributions from
non-planar diagrams,  while UV divergences arise only from planar
diagrams.  As we shall see later, A stringy limit,  i.e. a truncation 
to the planar diagrams,  
arises in a certain $\theta\rightarrow\infty$ limit.  This limit
is T-dual to a conventional 't Hooft $N \rightarrow \infty$ limit 
with fixed $m/N$,  where $m$ is the magnetic flux on a 
torus of fixed size. 

The usual correspondence between a $1/N$ expansion and a genus 
expansion arises when the Lagrangian is of 
the form $\frac{N}{\lambda} tr f(M)$, where the elements of the
$N \times N$ matrix $M$ are independent.  Let us now consider the 
action of $U(N)$
Yang Mills theory with a non-zero magnetic flux $m$ satisfying 
$gcd(N,m) =1$.  The gauge fields satisfy the boundary conditions
(\ref{bcs}).  The general solution of (\ref{bcs}) is
  \bea
{\tilde A}^{\mu}_{ij} = \sum_{\bf r} a^{\mu}_{\bf r} (Q^{br_1}P^{r_2})_{ij} 
        e^{i\pi \frac{b}{N} r_1r_2} 
        e^{i{\bf r} \cdot {\bf x}/(NR^{\prime})}.
\eea
Written in this form, ${\tilde A}$ does not have $N^2$ independent matrix 
elements for a given momentum.  There is an apparent reduction in 
the number of degrees of freedom from $N^2$ to $1$.  This can be understood
in terms of the duality relating the theory to non-commutative QED.  
Nevertheless, for sufficiently large momentum (and short distances)
the boundary conditions corresponding to the flux $m$ should have 
no effect on the physics.  This can be seen rather explicitly 
in the following way.  Roughly speaking,  one can exchange fractional
momentum modes for matrix degrees of freedom.  To do so,  one writes 
${\bf r} = N{\bf k} + {\bf l}$ with ${\bf l}$ defined modulo $N$, 
and defines a new ``gauge field''  $\cal A$ as follows;
\bea
{\cal A}^{\mu}_{ij} \equiv \sum_{\bf k} a^{\mu}_{{\bf k}N + {\bf l}} 
        (Q^{bl_1}P^{l_2})_{ij} 
                e^{i\pi \frac{b}{N} r_1r_2} 
                        e^{i{\bf k} \cdot {\bf x}/(R^{\prime})}.
\eea  
Since the $P$ and $Q$ matrices are complete,  one can write
\bea
{\cal A}^{\mu}_{ij} = \sum_{\bf k} f^{\mu}_{ij}({\bf k})
        e^{i{\bf k} \cdot {\bf x}/R^{\prime}}
\eea
with no constraints on $f^{\mu}_{ij}({\bf k})$.
If one considers a Yang-Mills action with the gauge field ${\cal A}$,
the usual 't Hooft large $N$ expansion applies.  
In momentum space, the difference between 
the Yang Mills action with the gauge field ${\cal A}$ or  ${\tilde A}$ lies in
terms proportional to $(k+j/N)/R^{\prime}$ as opposed to $k/R^{\prime}$.
At very
large $k$, i.e. $k >> j/N$,  there is no difference, as expected.  
However if one 
scales the radius of the torus as $R^{\prime} \sim 1/\sqrt{N}$,  then 
the modes that do not decouple as $N\rightarrow\infty$ have finite 
$(k+j/N)/R^{\prime}$, which is impossible if $k >> j/N$.
Thus the action is not of the form which
gives a genus expansion in $1/N$.

\section{An open string theory from $\theta\rightarrow\infty$.}

It was suggested in \cite{MRS} that the 
$\theta \rightarrow \infty$ limit of non-commutative field theory
gives rise to a string theory.  The reasons for this observation 
were as follows.  The feynman
graphs of a non-commutative theory can be written in a double line
notation,  just as in a theory with matrix indicies.  One can then
organize graphs acording to their genus.  In the large $\theta$ limit
the non-planar graphs are surpressed due to very oscillatory phases
in the loop integrations. The genus expansion is presumably an 
$l_s/\theta$ expansion where $l_s$ is an appropriate length
scale.

For a gauge theory,  the physical interpretation of a string theory 
arising in the $\theta\rightarrow\infty$ limit is not obvious
for the following reasons.      
Gauge theory on the non-commutative plane at finite $\theta$
has open string-like objects,  i.e. open Wilson lines.
However one would naively not expect such objects to exist
in the $\theta\rightarrow\infty$ limit,  since their length
in the non-commutative plane is $L = |\theta^{ij}p_j|$ which
diverges as $\theta\rightarrow\infty$.
The remaining gauge invariant operators are 
Wilson loops, i.e.
closed strings,  at zero momentum.  Thus in the 
$\theta\rightarrow\infty$ limit,  one might expect a closed string
theory in which space-time (in the non-commutative directions) has 
vanished and only global operators
survive.  Elaborating on a suggestion appearing in \cite{gt}, we 
will see below that there is a particular 
$\theta \rightarrow \infty$ limit of non-commutative gauge theory 
in which open strings do survive.  

One can obtain finite length open strings at infinite $\theta$ 
by compactifying the 
non-commutative plane on a torus and scaling $\theta$ and the radius 
to infinity simultaneously.  
Open Wilson lines on the torus are characterized by
a winding number (or electric flux)  in addition to their transverse
momentum (\cite{gt}). This is illustrated in figure 2.  
Gauge invariance requires the length of 
an open Wilson line propagating 
in, say,  the $1$ direction of the non-commutative $12$ torus to be
$L^2 = \theta^{21}p_1 + 2\pi R e_2$
where the integer $e_2$ is the electric flux.
At finite $p_1$, the first term in this expression for the length is 
infinite as
$\theta\rightarrow
\infty$.  On the other hand, the second term is also infinite
in the $R \rightarrow \infty$ if $e_2$ is non-zero.  
It is possible to
arrange the limit in such a way that the sum of these two terms
is finite.  

One way take such a limit is to set $\theta = -\frac{c}{N}2\pi R^2$ with 
$c=1$ and take 
$N \rightarrow \infty$ with $R/N = R^{\prime}$ fixed.
The resulting theory is dual to the large $N$ limit of a $U(N)$ Yang Mills 
theory with flux $m =N-1$, and {\it fixed} radius $R^{\prime}$.  Note that
this is now a conventional 't Hooft large $N$ limit,  with magnetic 
flux such that $m/N$ is fixed.  Consider a holonomy in the $U(N)$
theory which creates a state with electric flux $e_2^{\prime}$ 
and transverse momentum
$n_1^{\prime}$,  which are both held fixed in
the large $N$ limit.  Since $m/N$ and $R^{\prime}$ are  fixed in
this limit, 
the length of the open string in the non-commutative description 
\bea
L^2 = \theta^{21}p_1+ 2\pi R e_2 = 2\pi R^{\prime} e_2^{\prime}
\label{lengt}
\eea
is fixed, as is the transverse momentum 
\bea
p_1 = \frac{n_1}{R} = 
\frac{n_1^{\prime} - \frac{m}{N}e_2^{\prime}}{R^{\prime}}.
\label{momt}
\eea
where $m= N-1$.
Thus we obtain well defined open string states in the $\theta \sim N 
\rightarrow \infty$ limit with $R\sim N$. 

Since we are dealing with a standard 't Hooft limit in the 
dual $U(N)$ description, the genus expansion is an expansion in powers 
of $1/N$,  which plays the role of the string coupling $g_s$. 
In the non-commutative description, one would expect the genus 
expansion to be an expansion in $l_s^2/\theta$,
where $l_s$ is an appropriate length scale.  Since 
$1/N = {R^{\prime}}^2/\theta$,  we conclude that $l_s = R^{\prime}$.
Note that the string scale $l_s$ is related to the compactification scale $R$
by $l_s = \theta/R$.  
To summarize,
\bea
l_s = \frac{\theta}{R} \\
g_s = \frac{l_s^2}{\theta}
\eea
The length of open strings is quantized in units of 
$l_s$,  due to (\ref{lengt}),   
The momentum is also constrained 
via (\ref{momt})
to be $p_1 = L_2/(2\pi\theta) + j/l_s$, where $j$ is an 
integer\footnote{To see this write, $n_1^{\prime} = e_2^{\prime} + j$ in
(\ref{momt}).}. 

Strictly speaking, unless the open string theory is continuous in 
$\theta$, our discussion is only applicable to the case in 
which $\Theta = 1/N$.  
Continuity in $\theta$ should not be taken for 
granted when the space is compact\cite{CDS,gt,zg},  which is 
neccessarily the case if $g_s \ne 0$ and $l_s$ is finite.   

In some instances,  one can compute the correlator of
the dual open string states exactly at $g_s =0$ by using the $U(\infty)$ dual.
This was done in \cite{gt}, and we repeat the result here,
writing it in terms of non-commutative variables.  For pure
Euclidean NC-QED in two dimensions,  this correlator 
is defined by 
\bea
&& G_2(L_2,p_1) \equiv \nonumber \\ 
&& <\left(\frac{1}{R}\int d^2x P^* e^{i\int_{C} A_2} * e^{ip_1x^1}\right)
\left(\frac{1}{R}\int d^2y 
P\left(e^{-i\int_{C} A_2}\right) * e^{-ip_1y^1}\right)> \nonumber \\ 
\eea
where $C$ is the path from $(x^1,x^2)$ to $(x^1,x^2 + L_2)$.
At $g_s \rightarrow  0$ this can be computed in the $N\rightarrow\infty$
limit of the $U(N)$ description,  giving
\bea
G_2(L_2,p_1) =
e_2^{\prime}R^{\prime} 
\int_{-\infty}^{\infty} dy e^{i (n_1^{\prime}N - me_2^{\prime})y/(NR^{\prime})}
e^{-{g^{\prime}}^2 N e_2^{\prime} R^{\prime} |y|}. 
\eea
In terms of non-commutative variables,  this
may be rewritten as 
\bea 
G_2(L_2,p_1) =
\frac{2g^2L_2^2}{(g^2L_2)^2 + p_1^2}
\eea
where $g$ is the gauge coupling of NC-QED.
In the two dimensional case,  it is possible to compute 
the $1/N$ corrections although we shall not do so here.  

In the  ${\cal N} = 4$ version
of NC-QED in four dimensions,  
the equal time correlator of open
strings in the non-commutative plane can be computed at 
strong coupling using the dual $U(N)$ description and the 
AdS-CFT correspondence \cite{gt}.
In terms of non-commutative variables, one obtains
\bea
G_2(L_2,p_1) = 
L_2 \int_{-\infty}^{+\infty}dy 
e^{ip_1y}  
\exp\left(\frac{ L_2 (4\pi)^2(2g^2)^{1/2} }
         { \Gamma(\frac{1}{4})^4
                | y | } \right).
\eea

\begin{figure}
\begin{center}
\epsfig{file=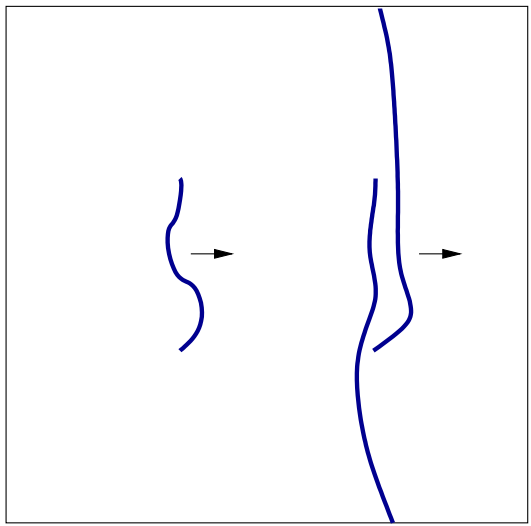, height=7cm,width=7cm}
\end{center}
\caption{Two open Wilson lines with the same transverse momentum.
Their length differs by $2\pi R n$ where $n$ is the difference in
electric fluxes.}
\label{figure2}
\end{figure}
  
\section*{Acknowledgements}

I am especially grateful to R. Jackiw and J. Troost for discussions
which lead to the present work.  I also profitted from conversations
with J. Feng, A. Leibovich, M. Schmaltz, W. Skiba,  Uwe-Jen Weise, 
and B. Zweibach. This work is supported in part by funds provided
by the U.S. Department of Energy (D.O.E) under cooperative research
agreement DE-FC02-94ER40818.

\end{document}